\documentclass[%
 aip,floatfix,
 amsmath,amssymb,
 reprint,%
]{revtex4-2}

\usepackage{graphicx}
\usepackage{dcolumn}
\usepackage{bm}
\usepackage[utf8]{inputenc}
\usepackage[T1]{fontenc}
\usepackage{mathptmx}
\usepackage{etoolbox}
\usepackage{multirow}
\usepackage{upgreek}
\usepackage{hyperref}

\makeatletter
\def\@email#1#2{%
 \endgroup
 \patchcmd{\titleblock@produce}
  {\frontmatter@RRAPformat}
  {\frontmatter@RRAPformat{\produce@RRAP{*#1\href{mailto:#2}{#2}}}\frontmatter@RRAPformat}
  {}{}
}%
\makeatother
\begin{document}
\newcommand{\Strath}{\affiliation{Department of Physics, SUPA, University of Strathclyde, Glasgow G4 0NG, United Kingdom}}
\newcommand{\Glasgow}{\affiliation{Kelvin Nanotechnology, University of Glasgow, Glasgow  G12 8LS, United Kingdom}}

\preprint{AIP/123-QED}

\title{Micro-machined deep silicon atomic vapor cells}
\author{S. Dyer}
\Strath
\author{P. F. Griffin}
\Strath
\author{A. S. Arnold}
\Strath
\author{F. Mirando}
\Glasgow
\author{D. P. Burt}
\Glasgow
\author{E. Riis}
\Strath
\author{J. P. McGilligan}
\email{james.mcgilligan@strath.ac.uk}
\Strath

\date{\today}

\begin{abstract}

Using a simple and cost-effective water jet process, silicon etch depth limitations are overcome to realize a $6\,$mm deep atomic vapor cell. While the minimum silicon feature size was limited to a $1.5\,$mm width in these first generation vapor cells, we successfully demonstrate a two-chamber geometry by including a $\sim$25~mm meandering channel between the alkali pill chamber and main interrogation chamber. We evaluate the impact of the channel conductance on the introduction of alkali vapor density during the pill activation process, and mitigate glass damage and pill contamination near the main chamber. Finally, we highlight the improved signal achievable in the $6\,$mm silicon cell compared to standard $2\,$mm path length silicon vapor cells.
\end{abstract}

\maketitle

\section{Introduction}

The micro-fabrication of atomic sensors has enabled a revolution in the field deployment and commercialization of chip-scale atomic sensors. \cite{Knappe2004,boudotmemscell, Hunter:18, kitchingmag} A critical component at the core of many of these chip-scale devices is the atomic vapor cell, consisting typically of a micro-electro-mechanical systems (MEMS) based glass-silicon-glass stack bonded under vacuum. \cite{moreland} Such MEMS cells have been widely explored for compact sensors, forming hermetic seals from both anodic and thermo-compression bonds to provide a mass-producible cell manufacturing  solution.  \cite{moreland,Kitching2018,karlencucu} While this now mature technology has enabled the transition to miniature components for atomic sensors, the methods of manufacturing silicon have typically limited the available optical path length to around $2\,$mm. These limitations arise due to dry etch phenomena \cite{DRIE} and challenges associated with long duration wet etch processes. \cite{mcgilliganreview} However, a longer optical path length would be beneficial for atomic sensors due to the increased atomic absorption and the associated enhancement in signal-to-noise ratio. 

To circumvent the short line-of-sight path length available in micro-fabricated cells, alternative light routing methods have been explored. A notable example is a wet-etched elongated silicon cell, with reflective walls at 54.7$^{\circ}$ from the wafer surface that direct the incident light through the long axis of the silicon cell, to achieve an optical path length of $7\,$mm. \cite{chutani} However, the additional complexity required to achieve optically smooth walls, as well as the inclusion of glass-etched grating structures, critically aligned to manipulate the angle of incidence for light routing parallel to the silicon surface, make this process costly and complex. While novel approaches to light routing in silicon cells may enable an increased optical path length, \cite{Nishino:21,nishino} the benefits of simple deep line-of-sight cells would also be valuable for chip-scale laser cooling, where the 6~mm silicon thickness could enable a larger optical overlap volume within the cell. \cite{Bregazzi}

Recent investigations of laser cooling in MEMS cells have reported a degradation of the achievable trapped atom number due to a limited silicon frame thickness, which restricts the optical overlap volume within the cell. \cite{Bregazzi} A viable solution for both hot and cold atom MEMS cells would be an increased optical path length in line-of-sight cells. 

Here we introduce a micro-machined silicon based vapor cell, manufactured with water jet cutting to achieve an optical path length of $6\,$mm in a simple line-of-sight geometry. An alkali pill deposited in the silicon frame provides a natural Rb density within the cell, with a meandering channel used to reduce material spray and glass damage in the main chamber of the cell. The step-change from conventional semiconductor fabrication processes to micro-machining for silicon-based vapor cell production greatly reduces the manufacturing costs by mitigating the need for photolithography and etching processes, while providing an improved optical path length to greatly increase the capabilities of chip-scale atomic sensors.

\section{Fabrication and set-up}

\begin{figure}[!t]
\centering
\includegraphics[width=1\columnwidth]{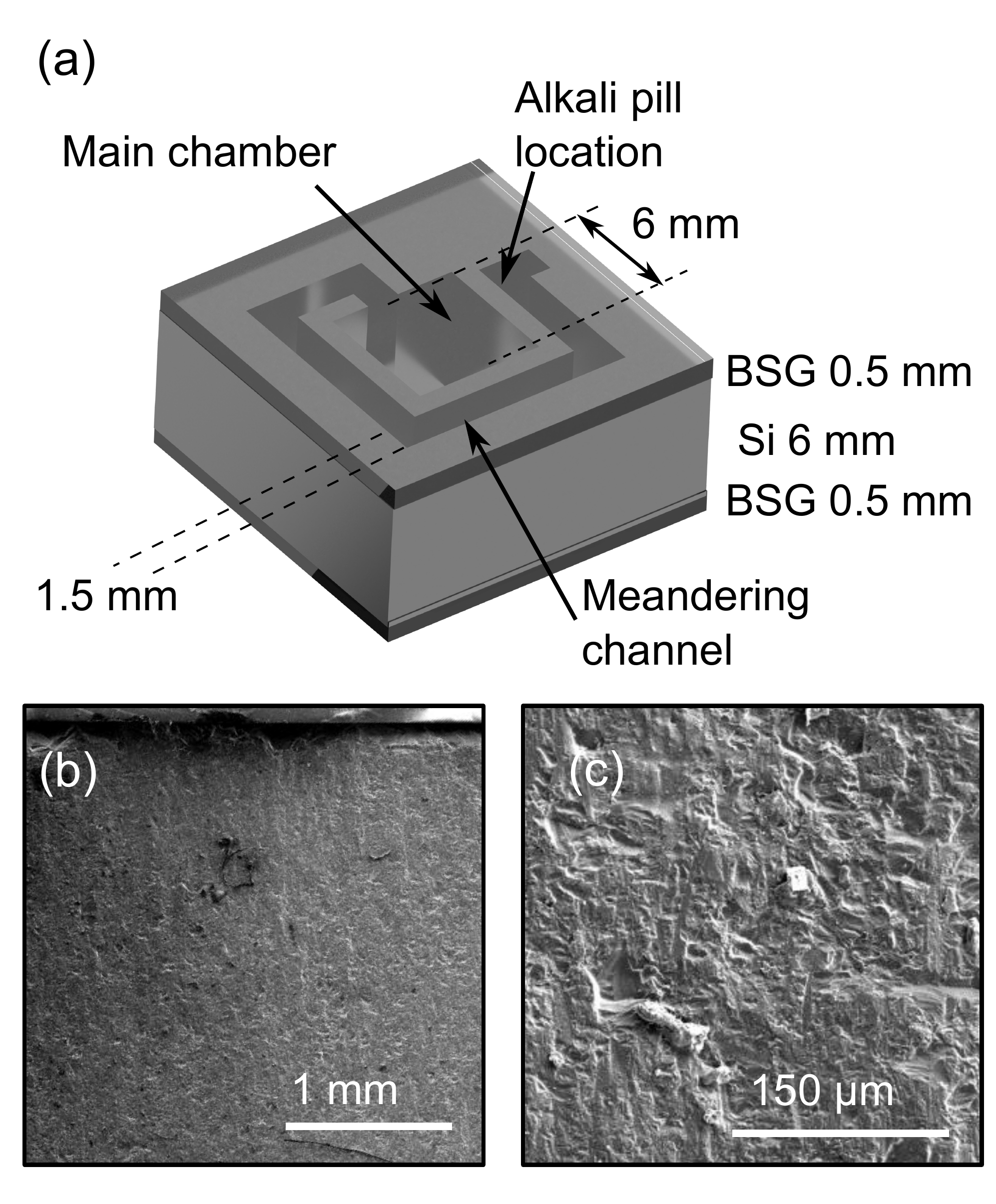}
\caption{\label{cell} (a): Illustration of the micro-machined silicon vapor cell. The cell is composed of a borosilicate glass (BSG) - Si - BSG stack, with total outer dimensions of ($13\,$mm$)^2\times 7\,$mm. 
The main cubic chamber has a $(6\,$mm$)^3$ volume, connected to an alkali micro-pill dispenser via a meandering channel with a cross-sectional area of $1.5\,$mm$\times 6\,$mm. (b)-(c): Scanning electron microscope (SEM) images of the micro-machined silicon surface.}
\end{figure}

The silicon frame of the cell is manufactured using a high pressure water jet cutting process. While this method is not limited to the depth restrictions placed upon micro-fabricated silicon, the minimum feature size is degraded. In this first generation cell, a minimum feature width of 1.5~mm was realized for the conductance channel between the pill source and main interrogation chamber. Non-line-of-sight conductance channels have been shown to be critical in reducing material spray and glass discoloration in the interrogation region of the vapor cell. \cite{karlencucu, Newman:19, boudotlineofsightchannel} Previously, non-line-of-sight channels have been dry etched with feature sizes on the order of $200\,\upmu$m$\times 200\,\upmu$m to greatly reduce the conductance into the main chamber, while permitting the presence of a suitable vapor density. \cite{Newman:19} Since the micro-machined channel's cross-sectional area is $\sim\times$220 larger than conventional dry etched channels \cite{Newman:19}, a long meandering path was machined between the pill location and main chamber, as shown in Fig.~\ref{cell} (a). The main cubic chamber was fabricated with dimensions $(6\,$mm$)^3$ 
to enable wider beams to be used for an improved signal-to-noise ratio in atomic spectroscopy.

Following the water jet cutting, the silicon inner wall surface roughness was evaluated with surface profilometry and scanning electron microscopy (SEM). SEM images shown in Fig.~\ref{cell} (b)-(c) were used to qualitatively evaluate the potential impact of wall roughness on the atomic vapor cell; as an increased surface area could consume more alkali content and hinder the homogeneity of applied wall coatings that may be required in later applications. The surface profilometry measurements determined an arithmetic average wall roughness of $R_a\,=\,5.9\,\upmu$m. While this wall roughness is comparable to recent work in additively manufactured vacuum chambers \cite{COOPER2021}, it remains orders of magnitude larger than the wall roughness achievable for DRIE, with short etch depths ($\leq$100~$\upmu$m) achieving $R_a=5\,$nm, \cite{DRIEsurface,DRIE} and deep etched silicon ($1.4\,$mm) demonstrating $R_a=300\,$nm \cite{CHUTANIdrie}. In future applications, chemical smoothing of the inner walls can be applied if the wall roughness is found to hinder the atomic vapor cell's long-term performance. \cite{CHUTANIdrie}

Following manufacturing, the silicon wafer was anodically bonded to $0.5\,$mm thick borosilicate glass wafers at both interfaces in a total background pressure of $\approx 10^{-5}\,$mbar. Prior to bonding, a commercially available alkali pill dispenser (SAES Rb/AMAX/Pill1-0.6) was deposited in the dedicated pill chamber of the silicon frame. The alkali pill was selected as the atomic source due to the relatively clean sourcing compared to azide and chloride compounds, \cite{Bopp_2020} enabling applications in wavelength referencing. \cite{boudotpill, newman2021highperformance} 
Post bonding, the wafer stack was diced into cells with total dimensions measuring $13\,$mm$\times 13\,$mm$\times 7\,$mm (L$\times$W$\times$H). To control the alkali vapor pressure, surface mounted resistors adhered to the front and back glass surfaces were used to heat the cell. The steady-state cell temperature was measured with a thermal camera.

A Yb fibre laser at $1070\,$nm was used for pill activation, and focused to fill the pill surface. A digital single-lens reflex camera was used to extract images of the cell window damage over time during the pill activation. A volume Bragg-reflector (VBR) laser at $780\,$nm was fibre-coupled and aligned through a polarizing beam-splitter (PBS) and the main chamber of the MEMS cell. The light was retro-reflected for in-situ `saturated absorption spectroscopy', \cite{Smith2004} with the photodiode using a $780\pm2\,$nm band-pass filter to avoid saturation from the Yb fibre laser during activation.

\section{Results}

\begin{figure}[!t]
\centering
\includegraphics[width=0.95\columnwidth]{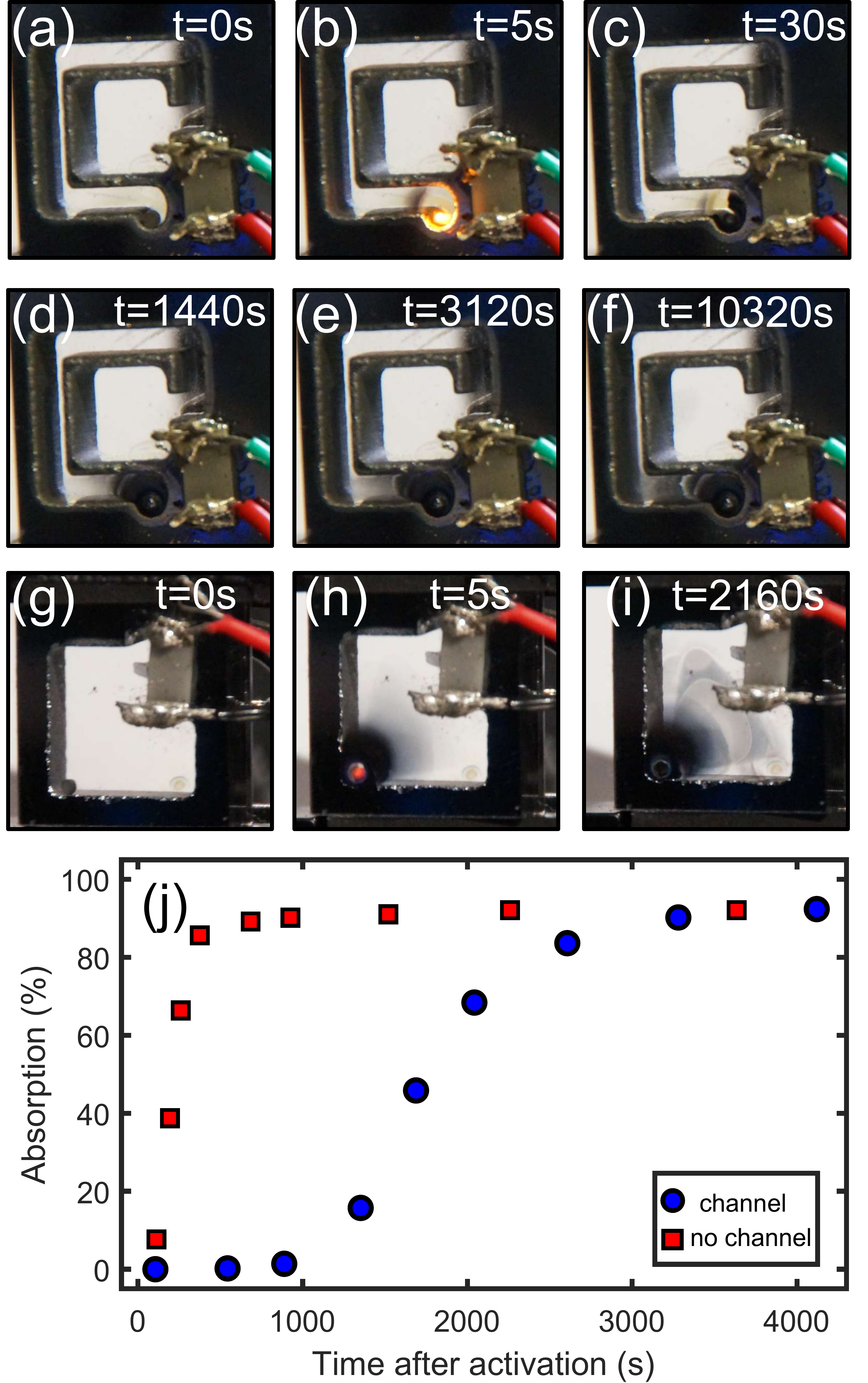}
\caption{\label{pill} Activation process of the alkali pill with laser heating: prior to activation (a); during the $10\,$s activation with $3\,$W of $1070\,$nm light (b); immediately after activation, where glass damage can be seen, as well as the presence of Rb on the glass surfaces (c). The migration of pill released material  after activation was then observed (d)-(f). Images (g)-(i) show a cell with no serpentine channel before, during, and after activation, respectively. 
(j); Simultaneously measured strongest D$_2$ transition absorption from the $^{85}$Rb $F=3$ ground state in the main chamber during the activation process for a channel and no-channel cell.}
\end{figure}

The alkali pill source in each individual cell was laser heated \cite{grifflaserheating} for $10\,$s with $3\,$W of $1070\,$nm light, focused to fill the $1\,$mm diameter of the pill surface. The activation process and subsequent material diffusion is highlighted in Fig.~\ref{pill} (a)-(f). In this demonstration, the cell was heated to 60$^{\circ}$C to drive alkali diffusion through the channel and into the main chamber. Images of a cell with no channel present during activation are shown in Fig.~\ref{pill} (g)-(i). The peak absorption evolution of the D$_2$ $^{85}$Rb $F=3$ ground state was simultaneously measured (Fig.~\ref{pill} (j)), from in-situ saturated absorption spectroscopy during the activation process of both cells types. For these measurements the laser intensity was 1.6$\,$mW/cm$^2$.
   
The cell prior to pill activation is seen at $t=0\,$s in (Fig.~\ref{pill} (a)), followed by a $10\,$s pulse of laser heating with $3\,$W of $1070\,$nm light, (shown at $t=5\,$s in Fig.~\ref{pill} (b)). While it is noted that the pill could be activated with lower laser powers, the chosen parameters ensured a reproducible activation between the cells under test. In the short time following activation, darkening of the glass is observed around the activation site, which progressively migrates through the conductance channel between $t=(30-10320)\,$s. Simultaneously, a silver spray of Rb metal can be seen migrating through the channel, with the growth of atomic absorption plateauing in the main chamber after $t\sim3000\,$s. Additionally, an identical cell was activated without heating during pill activation. In this scenario, the Rb accumulated near the pill activation site on the rough walls created by the mechanical micro-machining of the silicon frame. This was overcome by subsequent heating of the cell to again encourage migration into the main chamber.

In contrast, a comparable cell  without the meandering channel was fabricated, activated and observed in a similar manner (Fig.~\ref{pill} (g)-(i)). The absence of the channel between the pill and interrogation cell results in the activation process spraying material across the glass and reducing the cell optical access. Simultaneously, the absorption within the cell saturates much earlier at $t\sim400\,$s, as there is no diffusion time constant associated with reaching the cell interrogation region.

\begin{figure}[!t]
\centering
\includegraphics[width=\columnwidth]{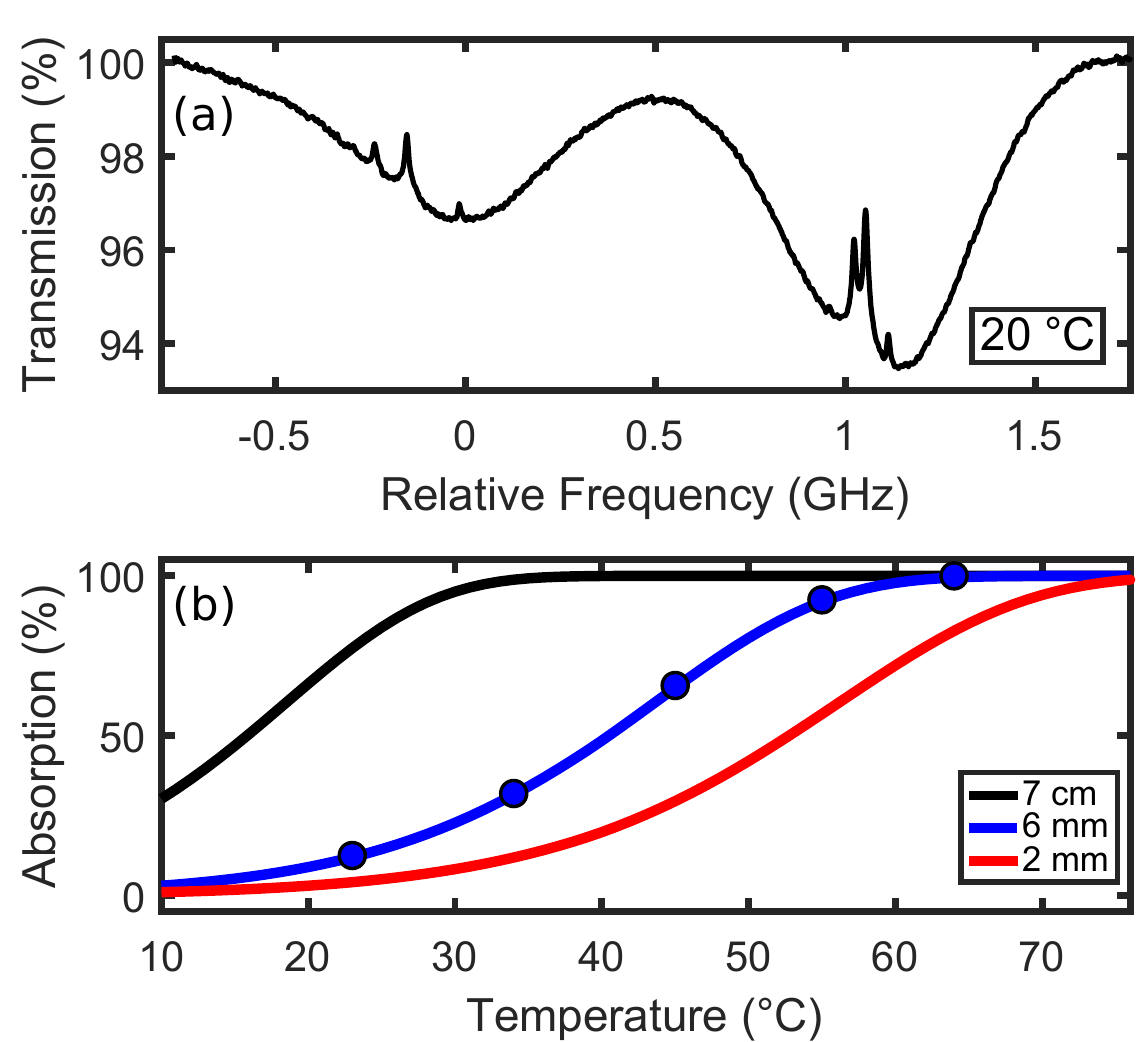}
\caption{\label{fig3} (a): Typical D$_2$ line 
saturated absorption spectrum (single-shot) for the $6\,$mm meandering micro-machined cell at room temperature. (b): Peak absorption from the $^{85}$Rb $F=3$ ground state Doppler-broadened profile as a function of temperature for retro-reflected vapor cells of $7\,$cm (black), $6\,$mm (blue) and $2\,$mm (red) path-lengths. Blue data points represent measurements taken with the meandering $6\,$mm vapor cell.}
\end{figure}

Following the activation process and measurement of alkali vapor density within the main cell, the cell absorption spectra were analysed, as shown in Fig.~\ref{fig3}. The spectra provided in Fig.~\ref{fig3} (a), was obtained with using a 1/e$^2$ beam waist of $\sim2\,$mm and peak intensity of $5\,$mW/cm$^2$. For this room temperature measurement, $\approx$7.5$\%$ peak D$_2$ absorption was measured from the $^{85}$Rb $F=3$ ground states. It is noted that the linewidth of the sub-Doppler features is $\sim\times1.4$ broader than comparable measurements in a glass reference cell over a range of incident intensities. While the cause of the broadening remains under investigation, the observation of a similar broadening in other pill based MEMS cells indicates that the cause could be from non-Rb content released during pill activation.

Fig.~\ref{fig3} (b) illustrates the peak absorption of the $^{85}$Rb $F=3$ ground state as a function of the measured cell temperature. The blue data points and curve show the measured and theoretically expected absorption at various temperature for a $6\,$mm cell length respectively. \cite{Siddons2008}. The $6\,$mm cell data are compared to theoretical curves for a typical $7\,$cm long glass blown cell (black), and a $2\,$mm path length micro-fabricated cell (red). For these measurements, a beam waist of $\sim2\,$mm and a peak intensity of 0.2$\,$mW/cm$^2$ were used. We observed good correlation between experimental and theoretical absorption.
For an operation temperature of 47$^{\circ}$C, the $6\,$mm cell is capable of achieving the same absorption level as a room temperature $7\,$cm cell, while the 2$\,$mm cell would require heating to 60$^{\circ}$C. Additionally, the optical path length of the deep silicon machined frame provides a sufficient signal-to-noise ratio for laser stabilization to the atomic reference cell at room temperature. 

In future work, we will evaluate the performance of the deep micro-machined cell as a wavelength reference for laser locking in a chip-scale laser cooling system. Additionally, the ability to fabricate 6~mm thick MEMS cells is directly transferable to the development of ultra-high-vacuum MEMS cells, where the increased vacuum volume would benefit the achievable optical overlap volume and trapped atom number. \cite{McGilligan2020,Bregazzi, mcgilliganSPIE}

\section{Conclusion}
\label{conclusion}

We have demonstrated the suitability of deep micro-machined silicon frames for the production of MEMS atomic vapor cells. While the first demonstration has a silicon feature width limitation of 1.5~mm, the inclusion of a meandering channel has been demonstrated to sufficiently restrict unwanted glass discoloration from the pill activation process outside of the main chamber, for unhindered optical interrogation. The increased optical path length was demonstrated in saturated absorption spectroscopy, with an atomic absorption significantly larger than is achievable in standard 2~mm MEMS vapor cells. The manufacturing simplicity and cost reduction of this technique makes this a highly attractive routine for macro-scale cell fabrication in atomic sensors where increased optical path lengths are desired.  

\begin{acknowledgments}
The authors would like to thank R.\ Boudot for useful conversations and careful reading of the manuscript. Additionally, we thank B.\ Lewis and A.\ Bregazzi for their reading of the manuscript. The authors acknowledge funding from Innovate UK project number 50133. J.~P.~M.\ gratefully acknowledges funding from a Royal Academy of Engineering Research Fellowship. 

\subsection{Conflict of interest}
The authors declare they have no competing interest. Approved for Public Release, Distribution Unlimited.
\end{acknowledgments}

\section*{Data Availability Statement}
The data that support the findings of this study are available from the corresponding author upon reasonable request.

\bibliography{Ref}

\end{document}